\documentstyle[12pt]{article}

\advance \textheight by \topskip
\def\ps#1{\raisebox{.2ex}{$\displaystyle
    \mathop{\psi}^{\scriptscriptstyle [#1]}$}{}}
\def\eps#1{\raisebox{.2ex}{$\displaystyle
    \mathop{\varepsilon}^{\scriptscriptstyle (#1)}$}{}}
\def\H#1{\raisebox{.2ex}{$\displaystyle
    \mathop{H}^{\scriptscriptstyle [#1]}$}{}}
\def\Lam#1{\raisebox{.2ex}{$\displaystyle
    \mathop{\Lambda}^{\scriptscriptstyle [#1]}$}{}}
\def\Om#1{\raisebox{.2ex}{$\displaystyle
    \mathop{\Omega}^{\scriptscriptstyle [#1]}$}{}}
\def\Ph#1{\raisebox{.2ex}{$\displaystyle
    \mathop{\Phi}^{\scriptscriptstyle [#1]}$}{}}
\def\u#1{\raisebox{.2ex}{$\displaystyle
    \mathop{u}^{\scriptscriptstyle [#1]}$}{}}
\def\A#1#2{{\mathop{A}\limits^{[#1]}_{[#2]}}{}}
\def\th#1{\raisebox{.2ex}{$\displaystyle
    \mathop{\theta}^{\scriptscriptstyle [#1]}$}{}}
\def\c#1{\raisebox{.2ex}{$\displaystyle
    \mathop{c}^{\scriptscriptstyle (#1)}$}{}}
\def\M#1{\raisebox{.2ex}{$\displaystyle
    \mathop{M}^{\scriptscriptstyle [#1]}$}{}}
\def\pp#1{\raisebox{.2ex}{$\displaystyle
    \mathop{\pi}^{\scriptscriptstyle [#1]}$}{}}
\def\et#1{\raisebox{.2ex}{$\displaystyle
    \mathop{\eta}^{\scriptscriptstyle [#1]}$}{}}
\def\p#1{\raisebox{.2ex}{$\displaystyle
    \mathop{\bar{p}}^{\scriptscriptstyle [#1]}$}{}}
\def\N#1{\raisebox{.2ex}{$\displaystyle
    \mathop{N}^{\scriptscriptstyle [#1]}$}{}}

\makeatletter
\@addtoreset{equation}{section}

\makeatother

\begin{document}

\begin{titlepage}
\hbox to \hsize{\hfil hep-th/9412134}
\hbox to \hsize{\hfil INR 867/94}
\hbox to \hsize{\hfil August, 1994}
\vfill
\large \bf
\begin{center}
BRST FORMALISM FOR SYSTEMS WITH HIGHER \\
ORDER DERIVATIVES OF GAUGE PARAMETERS
\end{center}
\vskip1cm
\normalsize
\begin{center}
{\bf Kh. S. Nirov\footnote{E--mail: nirov@ms2.inr.ac.ru}}\\
{\small Institute for Nuclear Research of the Russian Academy of
Sciences} \\
{\small 60th October Anniversary prospect 7a, 117312 Moscow, Russia}
\end{center}
\vskip2cm
\begin{abstract}
\noindent
For a wide class of mechanical systems, invariant under gauge
transformations with higher (arbitrary) order time derivatives
of gauge parameters, the equivalence of Lagrangian and Hamiltonian
BRST formalisms is proved. It is shown that the Ostrogradsky
formalism establishes the natural rules to relate the BFV ghost
canonical pairs with the ghosts and antighosts introduced by the
Lagrangian approach. Explicit relation between corresponding
gauge-fixing terms is obtained.
\end{abstract}
\vfill
\end{titlepage}

\section{Introduction}

Gauge invariant systems are described by singular Lagrangians.
Quantizing such a theory it is desirable to keep an initial covariance
of the system.
Hence, to have a consistent state space in corresponding quantum
theories, it is necessary to modify an initial velocity phase space
of such systems on the classical level.
The most popular way to do it -- BRST formalism -- is based on the notion
of the BRST-symmetry \cite{BRST},
discovered first for Yang-Mills field theory.

At present there have been elaborated two different approaches to construct
an effective BRST invariant theory from the initial singular system. In the
first approach, called Lagrangian BRST formalism \cite{Lagr}, one
starts from a gauge invariant system and, using consequences of the
gauge invariance, constructs a nonsingular BRST invariant Lagrangian.
Another method (Hamiltonian BRST (BFV) formalism \cite{BFV,Hen}) is
based on Hamiltonian description of the constrained system.  The
question of correspondence between Lagrangian and Hamiltonian BRST
formalisms was considered in many papers \cite{Equiv,NR1,NR2,NR3} by
the use of various approaches.

As we know, the most interesting from the physical point of view systems
have a gauge symmetry under transformations depending on time derivatives
of gauge parameters only up to the first order. The equivalence of
Lagrangian and Hamiltonian BRST formalisms for such systems was proven in
Refs.\cite{NR1,NR2,NR3} in such a way, that relations between BRST charges,
BRST invariant Hamiltonians and gauge-fixing terms of two approaches were
explicitly established. In these works there was also obtained the form
of the constraints and structure functions which appear when one
rewrites the corresponding Lagrangian functions through Hamiltonian
variables.  Now it would be interesting to perform the same analysis
for gauge invariant systems, whose symmetry transformations depend
arbitrarily on gauge parameters (we mean an arbitrary order of time
derivatives of infinitesimal gauge parameters).
Actually, within the framework of Lagrangian BRST formalism infinitesimal
parameters of gauge transformations are replaced by ghost variables. Then,
we shall get an effective BRST invariant system with higher order derivatives
because of nonsingular Lagrangian is constructed with the help of terms,
including BRST transformations of the velocity phase space coordinates. But
the BFV prescription introduces the ghosts as additional canonical variables,
simply associating them with the constraints.  Thus, there arises one
more question: what rules relate the ghosts of Lagrangian
and Hamiltonian BRST formalisms? Remember that the same question had also
appeared in the paper by I.V. Tyutin with collaborators
(see Ref.\cite{Equiv}). We will be convinced that within the framework
of our analysis the corresponding problem is consistently solved.

In Ref.\cite{N} Hamiltonian description for gauge invariant systems, having
the symmetry transformations with arbitrary dependence on gauge
parameters, has been constructed. In the same paper the explicit form of the
corresponding constraint algebra was obtained. We shall use the results of
Ref.\cite{N}.

This paper is organized in the spirit of \cite{NR2}. In Section 1 we
construct Lagrangian BRST formalism for gauge invariant systems with
arbitrary dependence on gauge parameters. By this we consider the
gauge transformations, depending only on the velocity phase space
coordinates.  Section 2 is devoted to Hamiltonian (BFV) BRST formalism
for considered mechanical systems. In Section 3 we prove the
equivalence between Lagrangian and Hamiltonian BRST formalisms and
present explicit connection of gauge-fixing terms of the two
approaches.

In this paper we restrict us to the initial gauge invariant systems
being bosonic, but the generalization of our results to the
case of mechanical systems, described by both even and odd variables,
is not a difficult problem \cite{DeW,PRR}.

We assume the summation over repeated indexes, and all the partial
derivatives to be the left partial derivatives.

\section{Lagrangian BRST formalism}

Let the velocity phase space \cite{Arn} be described by the set of
generalized coordinates $q^r$ and generalized velocities $\dot q^r$,
$r = 1,\ldots,R$. Consider on this space the mechanical system
given by Lagrangian $L(q,\dot q)$ having the symmetry under the gauge
transformations of the form
\begin{equation}
\delta_\varepsilon q^r = \sum_{k=0}^N {\eps{k}^\alpha \ps{N-k}^r_\alpha
(q,\dot q)}, \qquad \alpha = 1,\ldots,A,
\label{2.1}
\end{equation}
where $\varepsilon^\alpha$ are arbitrary infinitesimal functions of time,
and $N > 1$. Hence, we have
\begin{equation}
\delta_\varepsilon L = \frac{d}{dt} \Sigma_\varepsilon .
\label{2.2}
\end{equation}
As well as in Ref.\cite{N}, the following notations are used
in this paper:  integers within ordinary brackets (parentheses) over
characters display an order of time derivative of corresponding
functions, and all the integers within square brackets (both
subscripts and superscripts of characters) just mark the functions,
simply giving them numbering.

Equations of motion (Lagrange equations) of the system are differential
equations of the form
\begin{equation}
L_r(q,\dot q,\ddot q) \equiv W_{rs}(q,\dot q)\ddot q^s - R_r(q,\dot q) = 0,
\label{2.3}
\end{equation}
where
\begin{eqnarray}
R_r(q,\dot q) &=& \frac{\partial L(q,\dot q)}{\partial q^r} -
\dot q^s\frac{\partial^2 L(q,\dot q)}{\partial q^s\partial \dot q^r},
\label{2.4} \\
W_{rs}(q,\dot q) &=& \frac{\partial^2 L(q,\dot q)}{\partial \dot q^r
\dot q^s}. \label{2.5}
\end{eqnarray}
The matrix $W_{rs}$ is called the Hessian of the system.

{}From the symmetry equations (\ref{2.1}),(\ref{2.2}) we get the Noether
identities
\begin{equation}
\sum_{k=0}^N {(-1)^k \frac{d^k}{dt^k} \left(\ps{N-k}^r_\alpha L_r\right)}
= 0 , \label{2.6}
\end{equation}
which express the functional dependence of the Lagrange equations.
Using the Noether identities written in a more appropriate form
\begin{eqnarray}
\Lam{k+1}_\alpha &=& \ps{k}^r_\alpha R_r - \dot q^s
\frac{\partial \Lam{k}_\alpha}{\partial q^s} ,
\label{2.7} \\
\ps{k}^r_\alpha W_{rs} &=& - \frac{\partial \Lam{k}_\alpha}{\partial
\dot q^s} , \label{2.8}
\end{eqnarray}
where $k = 0,1,\ldots,N$ and $\Lam{0}_\alpha = \Lam{N+1}_\alpha \equiv 0$
by definition, we get for $\Sigma_\varepsilon$ the expression
\begin{equation}
\Sigma_\varepsilon = \delta_\varepsilon q^r \frac{\partial L}{\partial
\dot q^r} + \sum_{k=0}^{N-1} {\eps{k}^\alpha \Lam{N-k}_\alpha} .
\label{2.9}
\end{equation}

It follows from Noether identities (\ref{2.8}) that the Hessian of
the system is singular, and the Lagrange equations have no unique solution
for any initial values of $q^r(t)$ and $\dot q^r(t)$. We suppose the
null-vectors $\ps{0}^r_\alpha(q,\dot q)$, $\alpha = 1,\ldots,A$, of the
Hessian to be linearly independent, and any null-vectors of the
matrix $W_{rs}(q,\dot q)$ to be linear combinations of the vectors
$\ps{0}^r_\alpha$. Hence, we have
\begin{equation}
{\rm rank}\; W_{rs}(q,\dot q) = R - A , \qquad
{\rm rank}\; \ps{0}^r_\alpha(q,\dot q) = A  \label{2.10}
\end{equation}
for any values of $q^r$ and $\dot q^r$. Besides, let
the gauge transformations (\ref{2.1}) be nontrivial
for any choice of arbitrary functions $\varepsilon^\alpha(t)$ and
any trajectory of the system . It can
be shown that this condition is equivalent to the linear independence of
the set formed by the vectors $\ps{k}^r_\alpha$, $k = 0,1,\ldots,N$.

The Lagrange equations are solvable with respect to only
$R - A$ accelerations $\ddot q^r$, and, to determine the
evolution of the system, the Lagrangian constraints
\begin{equation}
\ps{0}^r_\alpha R_r = 0 , \label{2.11}
\end{equation}
must be kept. The latter follow directly from Eqs.(\ref{2.8}),(\ref{2.3}).
Such relations restrict the possible values of $q^r$ and $\dot q^r$ and
are called the primary Lagrangian constraints.

The stability condition for the primary Lagrangian constraints
$\Lam{1}_\alpha = \ps{0}^r_\alpha R_r$ gives rise to other
Lagrangian constraints of the system. From the Noether identities
we get that the complete set of the Lagrangian constraints of the
system is given by the relations
\begin{equation}
\Lam{k}_\alpha(q,\dot q) = 0, \qquad k = 1,\ldots,N.  \label{2.12}
\end{equation}

Suppose now that gauge transformations (\ref{2.1}) form a closed gauge
algebra. So, for any two sets of infinitesimal functions
$\varepsilon^\alpha_1(t)$ and $\varepsilon^\alpha_2(t)$ we have the
commutator of corresponding gauge transformations of type (\ref{2.1})
to be of the same type
\begin{equation}
\left[\delta_{\varepsilon_1} \,\ \delta_{\varepsilon_2}\right] q^r =
\delta_\varepsilon q^r ,
\label{2.13}
\end{equation}
where $\varepsilon^\alpha$ are, in general, some functions of
$\varepsilon^\alpha_1$, $\varepsilon^\alpha_2$ and the trajectory of the
system.

Taking into account the linear independence of the
vectors $\ps{k}^r_\alpha$, $k = 0,1,\ldots,N$, from
Eqs.(\ref{2.13}) and (\ref{2.1}) we obtain the relations of the
gauge algebra of the  system \cite{N}
\begin{eqnarray}
&&\ps{N-m+n}^s_\alpha \frac{\partial \ps{N-n}^r_\beta}{\partial q^s}
+ \left( \ps{N-m+n+1}^s_\alpha + \dot{\ps{N-m+n}^s_\alpha} \right)
\frac{\partial \ps{N-n}^r_\beta}{\partial \dot q^s}  \nonumber \\
&& \nonumber \\
&-& \ps{N-n}^s_\beta \frac{\partial \ps{N-m+n}^r_\alpha}{\partial q^s}
- \left( \ps{N-n+1}^s_\beta + \dot{\ps{N-n}^s_\beta} \right)
\frac{\partial \ps{N-m+n}^r_\alpha}{\partial \dot q^s} \nonumber \\
&& \nonumber \\
&=& \sum_{i=0}^N \sum_{j=0}^i
{\left( \begin{array}{c}
         i \\ j
 \end{array} \right)} \A{n-j}{m-i}^\gamma_{\alpha \beta}\,\ps{N-i}^r_\gamma
\nonumber \\
&& \nonumber \\
&+& \left[ \dot\A{n}{m}^\gamma_{\alpha \beta} \ps{1}^r_\gamma
+ \left( \ddot\A{n}{m}^\gamma_{\alpha \beta}
+ 2 \dot\A{n}{m-1}^\gamma_{\alpha \beta}
+ 2 \dot\A{n-1}{m-1}^\gamma_{\alpha \beta} \right) \ps{0}^r_\gamma \right] ,
\label{2.14}
\end{eqnarray}
where $n = 0,1,\ldots,N+1$; $m = 0,1,\ldots,2N+1$.
Here $\A{l}{k}^\gamma_{\alpha \beta}$ are some functions of the generalized
coordinates $q^r$, called the structure functions of the gauge algebras
(remember that for the case of $N = 1$ \cite{PR,NPR} the structure functions
of the corresponding gauge algebras, in general, depend on both generalized
coordinates and generalized velocities).

These functions satisfy the symmetry property
\begin{equation}
\A{l}{k}^\gamma_{\alpha \beta} = - \A{k-l}{k}^\gamma_{\beta \alpha} , \qquad
l \le k \le l + 1 , \label{2.15}
\end{equation}
and connect the infinitesimal parameters of gauge transformations in
(\ref{2.13}) by the relation
\begin{equation}
\varepsilon^\gamma = \sum_{k=0}^{N+1} \sum_{l=0}^k
\eps{k-l}^\alpha_1\,\,\eps{l}^\beta_2\,\,\A{l}{k}^\gamma_{\alpha \beta} .
\label{2.16}
\end{equation}

{}From (\ref{2.15}) we have that the only nonzero structure functions are
$\A{0}{0}^\gamma_{\alpha \beta}$,
$\A{0}{1}^\gamma_{\alpha \beta}$,
$\A{1}{1}^\gamma_{\alpha \beta}$, $\A{1}{2}^\gamma_{\alpha \beta}$.
Besides, as follows from Eq.(\ref{2.14}), for the case of
$N > 2$  all the structure functions are turned out to be constant,
thus the terms within the square brackets in the r.~h.~s. of
(\ref{2.14}) are equal to zero.

Finally, let us single out the following relations from the gauge
algebra
\begin{equation} \ps{0}^s_\alpha \frac{\partial
\ps{k}^r_\beta}{\partial \dot q^s} = \A{N-k}{N-k+1}^\gamma_{\alpha
\beta}\,\,\ps{0}^r_\gamma , \qquad k = 0,1,\ldots,N ,
\label{2.17}
\end{equation}
and note that for $k < N - 1$ the r.~h.~s. of Eq.(\ref{2.17}) is equal
to zero because of the above properties of the structure functions.

{}From the Jacobi identities for the gauge transformations (\ref{2.1})
\begin{equation}
\left( [ \delta_{\varepsilon_1}\,,\,[ \delta_{\varepsilon_2}\,,\,
\delta_{\varepsilon_3} ] ] + [ \delta_{\varepsilon_3}\,,\,[
\delta_{\varepsilon_1}\,,\,\delta_{\varepsilon_2} ] ]
+ [ \delta_{\varepsilon_2}\,,\,[ \delta_{\varepsilon_3}\,,\,
\delta_{\varepsilon_1} ] ] \right) q^r = 0 ,
\label{2.18}
\end{equation}
we get the relations
\begin{eqnarray}
&& \ps{N-l}^r_\alpha \frac{\partial}{\partial q^r}
\A{m}{n}^\varepsilon_{\beta \gamma} +
\ps{N-m}^r_\gamma \frac{\partial}{\partial q^r}
\A{n-m}{l-m+n}^\varepsilon_{\alpha \beta}
+ \ps{N-n+m}^r_\beta \frac{\partial}{\partial q^r}
\A{l}{l+m}^\varepsilon_{\gamma \alpha} \nonumber \\
&& \nonumber \\
&+& \sum_{i=0,1} \sum_{j=0}^i \sum_{k=0}^j
\Bigl( \A{i}{l+i}^\varepsilon_{\alpha \delta} \frac{d^{i-j}}{dt^{i-j}}
\A{m-k}{n-j}^\delta_{\beta \gamma} \nonumber \\
&& \nonumber \\
&+& \A{i}{m+i}^\varepsilon_{\gamma \delta} \frac{d^{i-j}}{dt^{i-j}}
\A{n-m-k}{l+n-m-j}^\delta_{\alpha \beta}
+ \A{i}{n-m+i}^\varepsilon_{\beta \delta} \frac{d^{i-j}}{dt^{i-j}}
\A{l-k}{l+m-j}^\delta_{\gamma \alpha} \Bigr) = 0 .
\label{2.19}
\end{eqnarray}
These relations, called the generalized Jacobi identities, give the
necessary and sufficient conditions for existence of the structure
functions of the gauge algebra.

To construct Lagrangian BRST formalism \cite{Lagr}, enlarge the
configuration space of the system, adding to (even) initial
generalized coordinates $q^r$ the set of odd variables $c^\alpha$,
$\alpha = 1,\ldots,A$, called the ghost variables, or simply the
ghosts, and odd variables $\bar{c}_\alpha$, called the antighosts. The
ghosts and antighosts are endowed with the ghost numbers, respectively
equal to $1$ and $-1$. Define the infinitesimal BRST transformations
for initial coordinates as follows \begin{equation} \delta_\lambda q^r
= \sum_{k=0}^N {\lambda \c{k}^\alpha \ps{N-k}^r_\alpha (q,\dot q)} ,
\label{2.20} \end{equation}
where $\lambda$ is the infinitesimal odd parameter.

Let $s$ be an odd vector field, connected with BRST transformations
(\ref{2.20}) by the relation
\begin{equation}
s( q^r ) = \sum_{k=0}^N { \c{k}^\alpha \ps{N-k}^r_\alpha(q,\dot q) } .
\label{2.21}
\end{equation}
{}From the nilpotency condition for BRST transformations we get the commutator
of the vector field $s$ with itself has to be equal to zero
\begin{equation}
[ s\,,\,s ] = 2 s^2 = 0 , \label{2.22}
\end{equation}
where the symbol $[ \;,\; ]$ means the generalized commutator of vector
fields \cite{DeW,PRR}.

Using the relations of the gauge algebra we obtain from condition
$s^2(q^r) = 0$ the expression for BRST transformation of the ghosts
\begin{equation}
s(c^\alpha) = - \frac{1}{2} \sum_{k=0}^{2} \sum_{l=0,1}
\c{k-l}^\beta\,\,\c{l}^\gamma\,\,\A{l}{k}^\alpha_{\beta \gamma} .
\label{2.23}
\end{equation}
{}From the Jacobi identities (\ref{2.19}) we get $s^2(c^\alpha) = 0$.

Introduce even auxiliary variables $b_\alpha$, $\alpha = 1,\ldots,A$, and
define BRST transformations for the antighosts as follows
\begin{equation}
s(\bar{c}_\alpha) = b_\alpha . \label{2.24}
\end{equation}
Supposing that
\begin{equation}
s(b_\alpha) = 0 , \label{2.25}
\end{equation}
we directly get $s^2(\bar{c}_\alpha) = 0$ and $s^2(b_\alpha) = 0$.

To remove the degeneracy of the initial gauge invariant Lagrangian,
one performs the ordinary BRST gauge--fixing procedure \cite{Lagr}.
To this end, introduce on the enlarged velocity phase space odd
function $F$, having the ghost number $-1$, and define a new Lagrangian
by the relation
\begin{equation}
L' = L + s(F) . \label{2.26}
\end{equation}
Choose $F$ in the most convenient form
\begin{equation}
F = \bar{c}_\alpha \chi^\alpha(q,\dot q) +
\frac{1}{2} \bar{c}_\alpha b_\alpha \gamma^{\alpha \beta} , \label{2.27}
\end{equation}
where $\gamma^{\alpha \beta}$ is some non--singular constant matrix, and
\begin{equation}
\chi^\alpha(q,\dot q) = \dot q^r \chi_r^\alpha(q) + \nu^\alpha(q) .
\label{2.28}
\end{equation}

Using BRST trans\-for\-ma\-tions of both even and odd va\-ri\-ab\-les,
we get from (\ref{2.26}) and (2.27) the ex\-pres\-sion
\begin{equation}
L' = L + b_\alpha \chi^\alpha + \frac{1}{2} b_\alpha b_\beta
\gamma^{\alpha \beta}
- \bar{c}_\alpha s(q^r) \frac{\partial \chi^\alpha}{\partial q^r} -
\bar{c}_\alpha s(\dot q^r) \frac{\partial \chi^\alpha}{\partial \dot q^r} .
\label{2.29}
\end{equation}
Taking into account the equations of motion for the auxiliary variables
$b_\alpha$, that are simple algebraic relations
\begin{equation}
b_\alpha = - \gamma_{\alpha \beta} \chi^\beta , \label{2.30}
\end{equation}
where $\gamma_{\alpha \delta} \gamma^{\delta \beta} = \delta^\beta_\alpha$,
we may rewrite the expression for $L'$ in the form
\begin{equation}
L'' = L - \frac{1}{2} \chi^\alpha \gamma_{\alpha \beta} \chi^\beta -
\bar{c}_\alpha s(q^r) \chi^\alpha_{;r} + \dot{\bar{c}}_\alpha s(q^r)
\frac{\partial \chi^\alpha}{\partial \dot q^r} - \frac{d}{dt}\left(
\bar{c}_\alpha s(q^r) \frac{\partial \chi^\alpha}{\partial \dot q^r} \right),
\label{2.31}
\end{equation}
where we use the notation $\chi^\alpha_{;r}$ for the variational derivative
$\frac{\partial \chi^\alpha}{\partial q^r} -
\frac{d}{dt}(\frac{\partial \chi^\alpha}{\partial \dot q^r})$.

Finally, removing from $L''$ the last term, which is a full time derivative,
we obtain the BRST invariant Lagrangian of the form
\begin{equation}
L_B = L - \frac{1}{2} \chi^\alpha \gamma_{\alpha \beta} \chi^\beta -
\bar{c}_\alpha s(q^r) \chi^\alpha_{;r} + \dot{\bar{c}}_\alpha s(q^r)
\frac{\partial \chi^\alpha}{\partial \dot q^r} .
\label{2.32}
\end{equation}

To obtain nondegenerate effective system, one needs to investigate the
super-Hessian, corresponding to the final Lagrangian $L_B$. One can
easily verify that the Lagrangian $L_B$ is nonsingular if and only if
the matrix
\begin{equation}
v^\alpha_\beta = \ps{0}^r_\beta \frac{\partial \chi^\alpha}{\partial
\dot q^r} = \ps{0}^r_\beta \chi_r^\alpha  \label{2.33}
\end{equation}
is nonsingular. We suppose that this is the case.

Let us obtain the BRST charge, corresponding to BRST symmetry of $L_B$.
The initial gauge invariant Lagrangian is BRST invariant, hence, from
the nilpotency of BRST transformations we get that $L'$ in
Eq.(\ref{2.29}) should be also BRST invariant. Note, that
\begin{equation}
s(L'') =
s(L') \vert_{b_\alpha = - \gamma_{\alpha \beta} \chi^\beta} ,
\label{2.34}
\end{equation}
and
\begin{equation}
s(L_B) = s(L'') + \frac{d}{dt} \left( s( \bar{c}_\alpha s(q^r)
\frac{\partial \chi^\alpha}{\partial \dot q^r}) \right) . \label{2.35}
\end{equation}
Hence, we have
\begin{equation}
s(L_B) = \frac{d}{dt} \Sigma_B , \label{2.36}
\end{equation}
where
\begin{equation}
\Sigma_B = s(q^r) \frac{\partial L}{\partial \dot q^r}
+ \sum_{k=0}^{N-1} {\c{k}^\alpha \Lam{N-k}_\alpha}
- s(q^r) \chi_r^\alpha \gamma_{\alpha \beta} \chi^\beta
+ \bar{c}_\alpha s(q^r) s(q^t)
\frac{\partial^2 \chi^\alpha}{\partial \dot q^r \partial q^t} .
\label{2.37}
\end{equation}

{}From Eqs.(\ref{2.36}), (\ref{2.37}) we get the expression for the
BRST charge
\begin{equation}
q_B = \Sigma_B - s(q^r) \frac{\partial L_B}{\partial \dot q^r}
- s(\bar{c}_\alpha) \frac{\partial L_B}{\partial \dot{\bar{c}}_\alpha}
\nonumber \\
- \sum_{k=0}^{N-1} \left( s(\c{k}^\alpha) \sum_{l=k}^{N-1} (-1)^{l-k}
\frac{d^{l-k}}{dt^{l-k}} \Bigl(\frac{\partial L_B}{\partial \c{l+1}^\alpha}
\Bigr) \right) .
\label{2.38}
\end{equation}

So, starting from the singular (gauge invariant) system, we have
constructed the effective nonsingular BRST invariant Lagrangian (\ref{2.32})
and obtained the expression for corresponding BRST charge (\ref{2.38}). Note
that, unlike the case of $N = 1$ \cite{NR1,NR2,NR3}, we have now the system
with higher order derivatives, since the Lagrangian $L_B$ contains time
derivatives of the ghosts up to $N$-th order. This circumstance forces
us to use further the Ostrogradsky formalism \cite{GT}.

\section{Hamiltonian BRST (BFV) formalism}

Consider a mechanical system, defined by the Hamiltonian $h$ and the
set of irreducible constraints of the first class $\varphi_a$
\cite{Dir}. We have the constraint algebra with respect to the
Poisson brackets of the form
\begin{eqnarray}
\{ h\,,\,\varphi_a \} &=& h^b_a \varphi_b , \label{3.1} \\
\{ \phi_a\,,\,\varphi_b \} &=& f^c_{ab} \varphi_c , \label{3.2}
\end{eqnarray}
where $h^b_a$ and $f^c_{ab}$ are some functions of the phase space
coordinates of the system. These functions called the structure
functions of the constraint algebra.

To construct a Hamiltonian BRST formalism \cite{BFV,Hen} for the
constrained system under consideration, let us enlarge the phase
space, adding to (even) initial coordinates, describing its points,
the set of odd variables $\theta^a$, $\pi_a$, associated with the
constraints $\varphi_a$. We put $\theta^a$ to be the ghost variables
with the ghost number $1$, whereas $\pi_a$ -- are canonically
conjugate to them generalized ghost momenta, having the ghost number
$-1$. The initial phase space coordinates have the ghost number equal
to zero. We set the Poisson brackets of the odd variables to be of
the form
\begin{equation}
\{ \theta^a\,,\,\theta^b \} = 0 , \qquad \{ \pi_a\,,\,\pi_b \} = 0 ,
\label{3.3}
\end{equation}
\begin{equation} \qquad \{ \pi_a\,,\,\theta^b \} = - \delta^b_a .
\label{3.4} \end{equation}

The principal ingredients of BFV formalism are the nilpotent (odd) BRST
charge with the ghost number equal to $1$, and the BRST invariant
Hamiltonian, which is an even function, having the ghost number equal
to zero. These functions are given on the extended phase space of even
and odd canonical variables, the general form of BRST charge being
represented by the following series
\begin{equation}
\Omega_B = \sum_{n \ge 0} \Om{n}_B = \sum_{n \ge 0}
\Om{n}^{b_1{\ldots}b_n}_{a_1{\ldots}a_{n+1}} \theta^{a_{n+1}}{\cdots}\,
\theta^{a_1} \pi_{b_n}{\cdots}\, \pi_{b_1} , \label{3.5}
\end{equation}
where
\begin{equation}
\Om{0}_{a_1} = \varphi_{a_1} , \label{3.6}
\end{equation}
and the quantities $\Om{n}^{b_1{\ldots}b_n}_{a_1{\ldots}a_{n+1}}$  $(n > 0)$
are determined by the nilpotency condition
\begin{equation}
\{ \Omega_B\,,\,\Omega_B \} = 0 . \label{3.7}
\end{equation}
The BRST invariant Hamiltonian may be written in the form
\begin{equation}
H_A = \sum_{n \ge 0} \H{n}_A = \sum_{n \ge 0}
\H{n}^{b_1{\ldots}b_n}_{a_1{\ldots}a_{n}} \theta^{a_n}{\cdots}\,
\theta^{a_1} \pi_{b_n}{\cdots}\, \pi_{b_1} . \label{3.8}
\end{equation}
Assuming, that
\begin{equation}
\H{0} = h , \label{3.9}
\end{equation}
we can find the quantities $\H{n}^{b_1{\ldots}b_n}_{a_1{\ldots}a_n}$
$(n > 0)$ from the BRST invariance condition for $H_A$
\begin{equation}
\{ \Omega_B\,,\,H_A \} = 0 . \label{3.10}
\end{equation}
The general theorem of existence of the higher order structure functions
$\Om{n}^{b_1{\ldots}b_n}_{a_1{\ldots}a_{n+1}}$  and
$\H{n}^{b_1{\ldots}b_n}_{a_1{\ldots}a_n}$ of BFV formalism has been proved
in Ref.\cite{Hen}. In particular, we have for $n = 1$
\begin{equation}
\Om{1}_B = - \frac{1}{2} f^c_{ab} \theta^b \theta^a \pi_c , \qquad
\H{1}_A = h^b_a \theta^a \pi_b . \label{3.11}
\end{equation}

{}From the nilpotency of the BRST charge we get that Eq.(\ref{3.10})
defines $H_A$ only up to BRST exact term. Hence, the general form
of the BRST invariant Hamiltonian is given by the expression
\begin{equation}
H_B = H_A - \{ \Omega_B\,,\,\Psi \} , \label{3.12}
\end{equation}
where $\Psi$ is an odd function, having the ghost number equal to $-1$.
Thus, the gauge--fixing procedure within the framework of Hamiltonian BRST
formalism consists of the choice of $\Psi$-function.

Hamiltonian formalism for the gauge invariant system, considered in the
previous Section, has been constructed in Ref.\cite{N}. In the same
paper the explicit relations of the constraint algebra were obtained.
Let us briefly recall the results of \cite{N}, that are necessary for
the further consideration.

Introduce $2R$--dimensional phase, the points of which are described by the
generalized coordinates $q^r$ and generalized momenta $p_r$, $r = 1,\ldots,R$.
Suppose the latter to be canonically conjugate pairs
\begin{equation}
\{ q^r\,,\,p_s \} = \delta^r_s , \label{3.13}
\end{equation}
and define a usual mapping of the velocity phase space to the phase space
as follows
\begin{equation}
p_r(q,\dot q) = \frac{\partial L(q,\dot q)}{\partial \dot q^r} .
\label{3.14}
\end{equation}
{}From the gauge invariance of the system we get that this mapping is
singular.  The image of the velocity phase space under the mapping,
given by (\ref{3.14}), is a $(2R - A)$--dimensional surface in the
phase space, the points of which may be described by the following
relations \begin{equation} \Ph{0}_\alpha (q,p) = 0 , \qquad \alpha =
1,\ldots,A , \label{3.15} \end{equation}
where the functions $\Ph{0}_\alpha$ are functionally independent.
Hence, we have introduced by Eq.(\ref{3.15}) the primary constraints
of the system \cite{Dir} and, respectively, the primary constraint
surface \cite{RS,PR,NR2,N}.

It can be shown that
\begin{equation}
\frac{\partial \Ph{0}_\alpha}{\partial p_r} (q,p(q,\dot q)) =
- \u{0}^\beta_\alpha (q,\dot q) \ps{0}^r_\beta (q,\dot q), \label{3.16}
\end{equation}
where $\u{0}^\beta_\alpha$ is nonsingular matrix. We choose this matrix to be
equal to the inverse one of the matrix $v^\beta_\alpha(q,\dot q)$, given by
Eq.(\ref{2.33}) :
\begin{equation}
v^\delta_\alpha \u{0}^\beta_\delta = \delta^\beta_\alpha , \qquad
v^\beta_\alpha = \chi^\beta_r \ps{0}_\alpha^r . \label{3.17}
\end{equation}

Let $F(q,p)$ be a function defined on the phase space. There
exists a function $f(q,\dot q)$ on the velocity phase space, such that
\begin{equation}
f(q,\dot q) = F(q,p(q,\dot q)) . \label{3.18}
\end{equation}
In this, the function  $f$ takes constant values at points of the
surfaces, given parametrically in the form \cite{RS}
\begin{eqnarray}
q^r(\tau) &=& q^r , \label{3.19} \\ \dot q^r(\tau) &=& \dot q^r +
\tau^\alpha \ps{0}^r_\alpha(q,\dot q) .  \label{3.20}
\end{eqnarray}
This fact may be easily expressed by the differential equations of the
form
\begin{equation}
\ps{0}^r_\alpha \frac{\partial f}{\partial \dot q^r} = 0 , \qquad
\alpha = 1,\ldots,A . \label{3.21}
\end{equation}

But for a given function $f(q,\dot q)$ on the velocity phase space,
there not always exists a corresponding function $F(q,p)$ on the phase
space, which is related to $f$ as follows
\begin{equation}
F(q,p(q,\dot q)) = f(q,\dot q) . \label{3.22}
\end{equation}
Eq.(\ref{3.21}) gives the necessary conditions for the existence of the
function $F(q,p)$.
We assume, following Ref.\cite{N}, that these relations are also
sufficient conditions for Eq.(\ref{3.22}) to be valid. It means that we
restrict us to the systems for which any point of the primary
constraint surface (\ref{3.15}) is the image of an unique surface of
the form (\ref{3.19}), (\ref{3.20}) \cite{RS}.

In this case, for any function $f(q,\dot q)$, which satisfies the
equalities (\ref{3.21}) one can find a function $F(q,p)$, connected
with $f$ by Eq.(\ref{3.22}) . We shall call such a function $f$ the
projectable to the primary constraint surface, or simply projectable,
and write
\begin{equation}
F \doteq f . \label{3.23}
\end{equation}
We have by definition
\begin{equation}
\Ph{0}_\alpha \doteq 0 , \label{3.24}
\end{equation}
hence, any function $F$ of the form
\begin{equation}
F = F_0 + F^\alpha \Ph{0}_\alpha , \label{3.25}
\end{equation}
where $F_0$ satisfies (\ref{3.23}) and $F^\alpha$ are
some arbitrary functions, fulfils the same equality (\ref{3.23}) as
well. Indeed, the relation (\ref{3.23}) determines the function $F$
only on the primary constraint surface and the solution of this
equation is defined up to a linear combination of the primary
constraints. Hence, the expression (\ref{3.25}) gives the general
solution of Eq.(\ref{3.23}). Note that the standard extension method
\cite{PR,NR3,N}, we shall use here, will consist in the way to fix the
specific form of functions $F_0$ and $F^\alpha$.

Introduce the energy function $E(q,\dot q)$:
\begin{equation}
E = \dot q^r \frac{\partial L}{\partial \dot q^r} - L . \label{3.26}
\end{equation}
This function is projectable to the primary constraint surface, hence
we can define the Hamiltonian of the system by the relation
\begin{equation}
H \doteq E . \label{3.27}
\end{equation}

Note that the Lagrangian constraints of the system, given by Eq.(\ref{2.12}),
satisfy the conditions (\ref{3.21}). This fact is a direct consequence of
the Noether identities. Hence, we may define the set of functions on the
phase space, corresponding to the Lagrangian constraints as follows
\begin{equation}
\Ph{k}_\alpha \doteq \Lam{k}_\alpha , \qquad k = 1,\ldots,N . \label{3.28}
\end{equation}
In Ref.\cite{N} it has been shown that the functions $\Ph{k}_\alpha$ are
the secondary Hamiltonian constraints of $k$-th stage. Note again that the
Eqs.(\ref{3.27}) and (\ref{3.28}) determine, respectively, the Hamiltonian
and the constraints of the system only on the primary constraint surface.
To get these functions and corresponding constraint algebra (with
respect to the Poisson brackets), it is necessary to define the way of
extension of functions from the primary constraint surface to the
total phase space.  It may be performed e.~g. within the framework of
the standard extension \cite{PR,N}. Remember that various extensions
differ from each other by a linear combination of the primary
constraints. Here we do not discuss the conditions for the
existence of the standard extension, assuming that they are valid, but
refer to Ref.\cite{PR,N}.

Function $F(q,p)$ is called the standard if it satisfies the relations
\begin{equation}
\chi^\alpha_r \frac{\partial F}{\partial p_r} = 0 , \label{3.29}
\end{equation}
where the vectors $\chi^\alpha_r(q)$ are dual to the vectors
$\u{0}^\beta_\alpha(q,\dot q) \ps{0}_\beta^r(q,\dot q)$ .
Let the Hamiltonian $H$ and constraints $\Ph{k}_\alpha$ be the standard
functions. Besides, we choose the so-called standard primary constraints,
satisfying the equalities
\begin{equation}
\Ph{0}_\alpha \doteq 0 , \qquad \frac{\partial \Ph{0}_\alpha}{\partial p_r}
= - ( \u{0}^\beta_\alpha \ps{0}^r_\beta )^0 , \label{3.30}
\end{equation}
where, and hereafter, the symbol $(f)^0$ denotes the standard Hamiltonian
analog for corresponding function $f$.
Then, the constraint algebra of the system under consideration in the
standard extension is given by the relations \cite{N}
\begin{eqnarray}
&&\{ \Ph{0}_\alpha\,,\,\Ph{0}_\beta \} \,=\,
\frac{\partial \Ph{0}_\alpha}{\partial p_r}\chi^\gamma_{rs}
\frac{\partial \Ph{0}_\beta}{\partial p_s}\;\Ph{0}_\gamma ,
\label{3.31} \\
&& \nonumber \\
&&\{ \Ph{k}_\alpha\,,\,\Ph{0}_\beta \} \,=\, \Bigl( \u{0}^\delta_\beta\,
\A{1}{N-k+1}^\gamma_{\alpha \delta} \Bigr)^0\;\Ph{1}_\gamma +
\frac{\partial \Ph{k}_\alpha}{\partial p_r}\chi^\gamma_{rs}
\frac{\partial \Ph{0}_\beta}{\partial p_s}\;\Ph{0}_\gamma ,
\label{3.32} \\
&& \nonumber \\
&&\{ \Ph{k}_\alpha\,,\,\Ph{l}_\beta \} \,=\, \Bigl( \u{k}^\delta_\alpha\,
\A{1}{N-l+1}^\gamma_{\beta \delta} - \u{l}^\delta_\beta\,
\A{1}{N-k+1}^\gamma_{\alpha \delta} + \dot q^r \frac{\partial}{\partial q^r}\,
\A{N-l}{2N-k-l}^\gamma_{\alpha \beta} \Bigr)^0\; \Ph{1}_\gamma \nonumber \\
&& \nonumber \\
&+& \sum_{i=0}^2 \sum_{j=0,1} {\left( \begin{array}{c}
                                     2N-k-l-i \\ N-l-j
                                    \end{array} \right)}
\A{j}{i}^\gamma_{\alpha \beta}\;\Ph{k+l-N+i}_\gamma
+ \frac{\partial \Ph{k}_\alpha}{\partial p_r}\chi^\gamma_{rs}
\frac{\partial \Ph{l}_\beta}{\partial p_s}\;\Ph{0}_\gamma ,
\label{3.33} \\
&& \nonumber \\
&&\{ H\,,\,\Ph{0}_\alpha \} \;=\, \bigl( \u{0}^\beta_\alpha \bigr)^0\;
\Ph{1}_\beta + \frac{\partial H}{\partial p_r}\chi^\beta_{rs}
\frac{\partial \Ph{0}_\alpha}{\partial p_s}\;\Ph{0}_\beta ,
\label{3.34} \\
&& \nonumber \\
&&\{ H\,,\,\Ph{k}_\alpha \} \;=\, \Ph{k+1}_\alpha -
\Bigl( \u{k}^\beta_\alpha + \mu^\delta \A{1}{N-k+1}^\beta_{\alpha \delta}
\Bigr)^0\; \Ph{1}_\beta + \frac{\partial H}{\partial p_r}\chi^\beta_{rs}
\frac{\partial \Ph{k}_\alpha}{\partial p_s}\; \Ph{0}_\beta ,
\label{3.35}
\end{eqnarray}
where $k, l = 1,\ldots,N > 1$, $i > N - k - l$, and we use the same
notations for $\mu^\alpha(q,\dot q)$, $\u{k}(q,\dot q)$ and
$\chi^\alpha_{rs}$ as in Ref.\cite{N}:
\begin{equation}
\mu^\alpha = \dot q^r \chi_r^\beta \u{0}^\alpha_\beta ,
\qquad \u{k}^\alpha_\beta = \ps{k}^r_\beta
\chi_r^\gamma \u{0}^\alpha_\gamma ,
\qquad \chi^\alpha_{rs} =
\frac{\partial \chi^\alpha_r}{\partial q^s} -
\frac{\partial \chi^\alpha_s}{\partial q^r} .
\label{3.36}
\end{equation}

Thus, we have the constraint system of the first class, and one can apply
to it the general BFV formalism, given in the beginning of this Section.
To this end, let us extend the initial phase space by adding to the
canonical pairs $q^r$, $p_r$ the set of odd ghost coordinates
$\et{k}^\alpha$ and ghost momenta $\pp{k}_\alpha$, $k = 0,\ldots,N$,
$\alpha = 1,\ldots,A$, having, the ghost numbers, respectively, $1$ and
$-1$. We suppose the non-zero Poisson brackets for the ghost variables
to be of the form
\begin{equation}
\{ \pp{k}_\alpha\,,\,\et{l}^\beta \} = - \delta^{kl} \delta^\beta_\alpha ,
\qquad k, l = 0,\ldots,N .
\label{3.37}
\end{equation}

The BRST charge $\Omega_B$ and the BRST invariant Hamiltonian $H_A$,
corresponding to the system with standard constraints $\Ph{k}_\alpha$ and
the standard Hamiltonian $H$, may be written in the form
\begin{eqnarray}
\Omega_B &=& \sum_{k=0}^N {\et{k}^\alpha \Ph{k}_\alpha} + \Delta{\Omega_B} ,
\label{3.38} \\
H_A &=& H + \Delta{H_A} , \label{3.39}
\end{eqnarray}
where $\Delta{\Omega_B}$ and $\Delta{H_A}$ consist of the terms of $n \ge 1$
according to Eqs.(\ref{3.5}), (\ref{3.8}).

Note that the BRST structure functions for $n = 1$ are given directly
by the structure functions of the constraint algebra
(\ref{3.31})--(\ref{3.36}) according to Eq.(\ref{3.11}). The BRST
structure functions of order $n = 2$ are constructed by using of the
Poisson brackets of the constraints and Hamiltonian with arbitrary
standard functions. The corresponding expressions for these Poisson
brackets has been calculated for the considered class of systems
$(N > 1)$ in Ref.\cite{N} (for the case of $N = 1$ see \cite{NR1,NR2,NR3}).

In this paper we shall not perform calculations of the higher order
BRST structure functions, but only prove in the next Section the
equivalence between Lagrangian and Hamiltonian BRST formalisms,
presented in the two previous Sections. Note that we will do it in the
spirit of Ref.\cite{NR2}.

\section{Relationship between Lagrangian and Hamiltonian approaches}

To compare the above formalisms, let us construct Hamiltonian
description for the system, given by the effective nonsingular BRST
invariant Lagrangian $L_B$ (\ref{2.32}). Recall that it corresponds
to the mechanical system with higher order derivatives. Hence one
should use the Ostrogradsky approach (see e.~g. \cite{GT}).
To this end, let us introduce the set of odd variables, putting
\begin{equation}
\th{k}^\alpha = \c{N-k}^\alpha , \qquad k = 1,\ldots,N . \label{4.1}
\end{equation}
Define the generalized momenta, corresponding to the system with the
Lagrangian $L_B$, as follows
\begin{eqnarray}
p_r &=& \frac{\partial L_B}{\partial \dot q^r} =
\frac{\partial L}{\partial \dot q^r} - \chi^\alpha_r \gamma_{\alpha \beta}
\chi^\beta - \bar{c}_\alpha \frac{\partial s(q^t)}{\partial \dot q^r}
\chi^\alpha_{;t} - \bar{c}_\alpha s(q^t) \chi^\alpha_{rt} +
\dot{\bar{c}}_\alpha \frac{\partial s(q^t)}{\partial \dot q^r} \chi^\alpha_t ,
\label{4.2} \\
&& \nonumber \\
p^\alpha &=& \frac{\partial L_B}{\partial \dot{\bar{c}}_\alpha} =
\sum_{k=0}^N \c{k}^\beta \ps{N-k}^r_\beta \chi^\alpha_r ,
\label{4.3} \\
&& \nonumber \\
\p{k}_\alpha &=& \sum_{l=1}^k (-1)^{k-l} \frac{d^{k-l}}{dt^{k-l}}
\left( \frac{\partial L_B}{\partial \c{N-l+1}^\alpha} \right) , \qquad
k = 1,\ldots,N. \label{4.4}
\end{eqnarray}
Remember that all the partial derivatives are understood as the left
partial derivatives \cite{DeW,PRR}.

{}From (\ref{4.3}), (\ref{4.4}) using Eqs.(\ref{3.17}), (\ref{3.36}) and
taking into account definition (\ref{4.1}), we get
\begin{eqnarray}
\dot{\bar{c}}_\alpha &=& - \bigl( \p{1}_\beta - \bar{c}_\gamma
\chi^\gamma_{;r} \ps{0}^r_\beta \bigr) \u{0}^\beta_\alpha ,
\label{4.5} \\
&& \nonumber \\
\c{N}^\alpha &=& p^\beta \u{0}^\alpha_\beta - \sum_{k=1}^N \th{k}^\beta
\u{k}^\alpha_\beta .
\label{4.6}
\end{eqnarray}

Let us introduce the projector
\begin{equation}
\Pi^r_s = \delta^r_s - \chi_s^\alpha \u{0}^\beta_\alpha \ps{0}^r_\beta ,
\qquad \Pi^t_s \Pi^r_t = \Pi^r_s , \label{4.7}
\end{equation}
and define for the singular Hessian $W_{rs}$ the corresponding pseudo-inverse
matrix $W^{rs}$, which is uniquely determined by the relations \cite{R}
\begin{equation}
W^{rt} W_{ts} = \Pi^r_s , \qquad W^{rs} \chi^\alpha_s = 0 . \label{4.8}
\end{equation}
Representing the BRST transformations of the generalized coordinates $q^r$ in
the form
\begin{equation}
s(q^r) = p^\beta\, \u{0}^\alpha_\beta\, \ps{0}^r_\alpha + \sum_{k=1}^N
\th{k}^\alpha \ps{k}^s_\alpha \Pi^r_s , \label{4.9}
\end{equation}
we may rewrite the expression for the generalized momenta $p_r$ as follows
\begin{equation}
p_r = \M{0}_r + \M{1}_r , \label{4.10}
\end{equation}
where we use the notations
\begin{eqnarray}
\M{0}_r &=& \frac{\partial L}{\partial \dot q^r} - \chi^\alpha_r
\gamma_{\alpha \beta} \chi^\beta , \label{4.11} \\
&& \nonumber \\
\M{1}_r &=& \left( - p^\alpha \frac{\partial
\u{0}^\beta_\alpha}{\partial \dot q^r}
+ \sum_{k=1}^N \th{k}^\alpha
\frac{\partial \u{k}^\beta_\alpha}{\partial \dot q^r}
\right) \p{1}_\beta + \sum_{k=1}^N \th{k}^\alpha
\frac{\partial (\ps{k}^s_\alpha \Pi^t_s)}{\partial \dot q^r}
\chi^\beta_{;t}\, \bar{c}_\beta \nonumber \\
&& \nonumber \\
&+& \left( p^\alpha\, \u{0}^\beta_\alpha\, \ps{0}^t_\beta + \sum_{k=1}^N
\th{k}^\alpha \ps{k}^s_\alpha \Pi^t_s \right) \chi^\gamma_{rt}\,
\bar{c}_\gamma .
\label{4.12}
\end{eqnarray}

{}From Eqs.(\ref{2.37}), (\ref{2.38}) using Eqs.(\ref{4.5}), (\ref{4.6})
and (\ref{4.10})--(\ref{4.12}) we get the following expression for
the BRST charge
\begin{equation}
q'_B = p^\alpha \gamma_{\alpha \beta} \chi^\beta +
\sum_{k=1}^N \th{k}^\alpha \Lam{k}_\alpha \nonumber \\ - s(q^r) \M{1}_r -
\sum_{k=1}^N s(\th{k}^\alpha) \p{k}_\alpha + \frac{1}{2} s(q^r) s(q^t)
\chi^\alpha_{rt}\, \bar{c}_\alpha , \label{4.13}
\end{equation}
where $s(q^r)$ is given by Eq.(\ref{4.9}). Hence, we have expressed the
BRST charge via generalized ghost coordinates and ghost momenta, and initial
even variables $q^r$, $\dot q^r$.

Now according to the Ostrogradsky formalism we introduce the energy
function which corresponds to the effective Lagrangian $L_B$ as follows
\begin{equation}
E_B = \dot q^r \frac{\partial L_B}{\partial \dot q^r} + \dot{\bar{c}}_\alpha
\frac{\partial L_B}{\partial \dot{\bar{c}}_\alpha} + \sum_{k=0}^{N-1}
\c{N-k}^\alpha \sum_{l=1}^{k+1} (-1)^{k+1-l} \frac{d^{k+1-l}}{dt^{k+1-l}}
\left( \frac{\partial L_B}{\partial \c{N-l+1}^\alpha} \right) - L_B .
\label{4.14}
\end{equation}
Using the gauge algebra relations (\ref{2.17}) and properties of the
projector $\Pi^r_s$, from Eqs.(\ref{2.32}) and (\ref{4.14}) we see
that the energy function $E_B$, expressed through the Hamiltonian ghost
variables (\ref{4.5}), (\ref{4.6}) and even coordinates of the initial
velocity phase space, has the form
\begin{eqnarray}
E'_B &=& E + \dot q^s \Pi^r_s \M{1}_r + p^\alpha\,
\u{0}^\beta_\alpha\, \p{1}_\beta + \sum_{k=1}^N \th{k}^\alpha \left(
\p{k+1}_\alpha - \Bigl(\u{k}^\beta_\alpha + \mu^\delta
\A{1}{N-k+1}^\beta_{\alpha \delta}\Bigr) \p{1}_\beta \right) \nonumber \\
&& \nonumber \\
&+& \left( \nu^\alpha - \frac{1}{2} \chi^\alpha \right)
\gamma_{\alpha \beta} \chi^\beta - \left( p^\alpha\,
\u{0}^\delta_\alpha\, \ps{0}^r_\delta + \sum_{k=1}^N \th{k}^\alpha
\ps{k}^s_\alpha \Pi^r_s \right) \frac{\partial \nu^\beta}{\partial
q^r}\, \bar{c}_\beta , \label{4.15}
\end{eqnarray}
where the functions $\u{k}^\alpha_\beta$ and $\mu^\alpha$ are given by
Eq.(\ref{3.36}), and $E$ is the energy function, corresponding to the
initial gauge invariant Lagrangian $L$.

Making use of the method, presented in Ref.\cite{NR2}, we may express the
generalized velocities $\dot q^r$ in the form of expansion over even powers
of the Hamiltonian ghost variables (the coefficients of such an expansion
will be functions of even canonical variables $q^r$, $p_r$).
In order to do this, denote the functions entering right-hand side
of (\ref{4.10})-(\ref{4.12}) by $M_r(q,\dot q,\theta,\pi)$:
\begin{equation}
M_r = \sum_{n=0,1} \M{n}_r , \label{4.16}
\end{equation}
where the functions $\M{0}_r$ do not depend on ghost variables $\theta^a$,
$\pi_a$, while the functions $\M{1}_r$ are quadratic in these variables.
Let us denote the functions, expressing the variables $\dot q^r$ via $q^r$,
$p_r$ and $\theta^a$, $\pi_a$ by $N^r(q,p,\theta,\pi)$. Thus, we have the
equality
\begin{equation}
M_r(q,N(q,p,\theta,\pi),\theta,\pi) = p_r , \label{4.17}
\end{equation}
that may be shortly written as
\begin{equation}
M_r(N) = p_r . \label{4.18}
\end{equation}
Represent the functions $N^r$ in the form
\begin{equation}
N^r = \sum_{n \ge 0} \N{n}^r , \label{4.19}
\end{equation}
where the functions $\N{n}^r$ have the degree $2n$ in ghost variables
$\theta^a$ and $\pi_a$. Expanding (\ref{4.18}) over degrees of the ghost
variables, we get
\begin{equation}
\sum_{n=0,1} \sum_{k \ge 0} \left( \frac{1}{k!}
\frac{\partial^k\M{n}(\N{0})}{\partial\dot q^{s_1}\ldots\partial\dot q^{s_k}}
\sum_{l_1,\ldots,l_k \ge 1} \N{l_1}^{s_1} \cdots \N{l_k}^{s_k} \right)
= p_r . \label{4.20}
\end{equation}
In particular, we have
\begin{eqnarray}
&& \M{0}_r(\N{0}) = p_r , \label{4.21} \\
&&\frac{\partial \M{0}_r(\N{0})}{\partial \dot q^{s_1}} \N{1}^{s_1} +
\M{1}_r(\N{0}) = 0 . \label{4.22}
\end{eqnarray}
Recall that the functions $\N{0}^r$ have the following important property
\cite{NR2}. Let a function $f(q,\dot q)$ be projectable, so that
\begin{equation}
\ps{0}^r_\alpha \frac{\partial f}{\partial \dot q^r} = 0 . \label{4.23}
\end{equation}
Consider the function $F$ connected with $f$ by the relation
\begin{equation}
F = f(\N{0}) . \label{4.24}
\end{equation}
{}From (\ref{4.21}) it follows that
\begin{equation}
f = F(\M{0}) . \label{4.25}
\end{equation}
Using this equality, it is easy to show that
\begin{equation}
\chi_r^\alpha \frac{\partial F}{\partial p_r} = 0 . \label{4.26}
\end{equation}
Hence, the function $F$ is a standard function.
For any standard function $F$ we have
\begin{equation}
F(\M{0}) = F(\partial L/\partial \dot q^r) . \label{4.27}
\end{equation}
Therefore, if the function $f$ satisfies the conditions (\ref{4.23}), then
\begin{equation}
f(\N{0}) = f^0 . \label{4.28}
\end{equation}
Introduce the notation
\begin{equation}
G_{rs} = \frac{\partial \M{0}_r}{\partial \dot q^s} . \label{4.29}
\end{equation}
{}From (\ref{4.11}) it follows that
\begin{equation}
G_{rs} = W_{rs} - \chi_r^\alpha \gamma_{\alpha \beta} \chi_s^\beta .
\label{4.30}
\end{equation}
It is easy to check that the matrix $G_{rs}$ is nonsingular.
Actually, we have
\begin{equation}
G^{rt} G_{ts} = \delta^r_s , \qquad G^{rs} = W^{rs} - \ps{0}^r_\gamma
\u{0}^\gamma_\alpha \gamma^{\alpha \beta} \u{0}^\delta_\beta
\ps{0}^s_\delta . \label{4.31}
\end{equation}
{}From (\ref{4.22}) we now get the equality
\begin{equation}
\N{1}^r(\M{0}) = - G^{rs} \M{1}_s . \label{4.32}
\end{equation}

Finally, using Eqs.(\ref{4.28}), (\ref{4.32}) from (\ref{4.13}) we get
the following expression for the BRST charge
\begin{eqnarray}
Q_B &=& p^\alpha \Ph{0}_\alpha + \sum_{k=1}^N \th{k}^\alpha \Ph{k}_\alpha
+ \sum_{k=1}^N \Bigl( \u{0}^\delta_\beta \A{1}{N-k+1}^\gamma_{\alpha \delta}
\Bigr)^0 \;p^\beta\,\th{k}^\alpha\,\p{1}_\gamma \nonumber \\
&& \nonumber \\
&+& \frac{1}{2} \sum_{k,l=1}^N \Bigl( \u{k}^\delta_\alpha\,
\A{1}{N-l+1}^\gamma_{\beta \delta} - \u{l}^\delta_\beta\,
\A{1}{N-k+1}^\gamma_{\alpha \delta} + \dot q^r \frac{\partial}{\partial q^r}\,
\A{N-l}{2N-k-l}^\gamma_{\alpha \beta} \Bigr)^0\; \th{k}^\alpha\,\th{l}^\beta\,
\p{1}_\gamma \nonumber \\
&& \nonumber \\
&+& \frac{1}{2} \sum_{k,l=1}^N \sum_{i=0}^2 \sum_{j=0,1}
                                 {\left( \begin{array}{c}
                                         2N-k-l-i \\ N-l-j
                                         \end{array} \right)}
\A{j}{i}^\gamma_{\alpha \beta}\;\th{k}^\alpha\, \th{l}^\beta\;
\p{k+l-N+i}_\gamma \nonumber \\
&& \nonumber \\
&+& \frac{1}{2}\left(
\frac{\partial \Ph{0}_\alpha}{\partial p_r}\chi^\gamma_{rs}
\frac{\partial \Ph{0}_\beta}{\partial p_s}\right)
p^\alpha\, p^\beta\, \bar{c}_\gamma
+ \sum_{k=1}^N \left(\frac{\partial \Ph{k}_\alpha}{\partial p_r}
\chi^\gamma_{rs}
\frac{\partial \Ph{0}_\beta}{\partial p_s}\right)
\th{k}\, p^\beta\, \bar{c}_\gamma \nonumber \\
&& \nonumber \\
&+& \frac{1}{2}\sum_{k,l=1}^N
\left(\frac{\partial \Ph{k}_\alpha}{\partial p_r}\chi^\gamma_{rs}
\frac{\partial \Ph{l}_\beta}{\partial p_s}\right) \th{k}^\alpha\,
\th{l}^\beta\, \bar{c}_\gamma  + \Delta{Q_B} , \label{4.33}
\end{eqnarray}
where $\Delta{Q_B}$ contains the terms having more than cubic powers
in the ghost variables. Note that all the constraints and higher
order structure functions in Eq.(\ref{4.33}) turn out to be the
standard functions. Besides, we have
\begin{equation}
\{ Q_B\,,\,Q_B \} = 0 . \label{4.34}
\end{equation}
Indeed, the vector field $s$,
which defines the BRST transformations, satisfies the nilpotency
condition (\ref{2.22}), at least on the equations of
motion, following from $L_B$. BRST charge $Q_B$ is nothing but the
Hamiltonian analog of the vector field $s$. Hence, the Poisson
brackets $\{ Q_B\,,\,Q_B \}$ have to be constant. Since the ghost
number of this constant is equal to $2$, we get it to be zero.

Now it is quite natural to identify the ghost variables of the
previous Section with those we have in this one by the following rules
\begin{eqnarray}
p^\alpha = \et{0}^\alpha , &\qquad& \th{k}^\alpha = \et{k}^\alpha ,
\label{4.35} \\
\bar{c}_\alpha = \pp{0}_\alpha , &\qquad& \p{k}_\alpha = \pp{k}_\alpha ,
\label{4.36}
\end{eqnarray}
for any $k = 1,\ldots,N$. From Eqs.(\ref{3.38}), (\ref{4.33}) we now
see that, using the arbitrariness in definition of the constraints,
one can always get
\begin{equation}
Q_B = \Omega_B . \label{4.37}
\end{equation}

To proceed to the Hamiltonian, consider the following odd function,
having the ghost number equal to $-1$,
\begin{equation}
\psi = \bar{c}_\alpha \left( \nu^\alpha - \frac{1}{2} \chi^\alpha \right) .
\label{4.38}
\end{equation}
Taking into account the equations of motion, following from $L_B$, we obtain
the expression for BRST transformation of this function
\begin{equation}
s(\psi) = - \left( \nu^\alpha - \frac{1}{2} \chi^\alpha \right)
\gamma_{\alpha \beta} \chi^\beta +
s(q^r) \frac{\partial \nu^\alpha}{\partial q^r} \bar{c}_\alpha . \label{4.39}
\end{equation}
Using the above reasoning, we get the relation
\begin{equation}
\{ \Psi\,,\,Q_B \} (M) = s(\psi) , \label{4.40}
\end{equation}
where $\Psi$ is the Hamiltonian analog of odd function $\psi$.
Explicitly, $\Psi$-function is given by the expression
\begin{eqnarray}
\Psi &=& \bar{c}_\alpha \left( \nu^\alpha - \frac{1}{2}
\gamma^{\alpha \beta} \left( \Ph{0}_\beta - \sum_{k=1}^N \th{k}^\delta\,
\left(\u{0}^\gamma_\beta \A{1}{N-k+1}^\varepsilon_{\delta \gamma}\right)^0
\p{1}_\varepsilon \right) \right)
\nonumber \\
&& \nonumber \\
&+& \frac{1}{2} \bar{c}_\alpha \gamma^{\alpha \beta} \left( \left( p^\delta
\left( \u{0}^\varepsilon_\delta\; \ps{0}^t_\varepsilon\;
\ps{0}^r_\beta \right)^0 + \sum_{k=1}^N \th{k}^\delta \,\left(\ps{k}^s_\delta
\Pi^t_s \ps{0}^r_\beta \right)^0 \right)
\chi^\gamma_{rt}\, \bar{c}_\gamma \right)
+ \Delta{\Psi} , \label{4.41}
\end{eqnarray}
where $\Delta{\Psi}$ consists of the terms of more than cubic powers in the
ghost variables. Note that for the primary constraints, which are linear in
generalized momenta $p_r$ this term $\Delta{\Psi}$ is equal to zero.

Denote the Hamiltonian, corresponding to the effective Lagrangian $L_B$, by
$H_B$, and consider the function $H_A$, connected with $H_B$ by the
relation
\begin{equation}
H_B = H_A - \{ \Psi\,,\,Q_B \} .
\label{4.42}
\end{equation}
Since the BRST transformations are the symmetry transformations for
the Lagrangian $L_B$, we see that the relation
\begin{equation}
\{ Q_B\,,\,H_B \} = 0  \label{4.43}
\end{equation}
is indeed valid. Hence, because of nilpotency of the BRST charge
$Q_B$, the Hamiltonian $H_A$ is also BRST invariant
\begin{equation}
\{ Q_B\,,\,H_A \} = 0 . \label{4.44}
\end{equation}
Now, using the technique of Ref.\cite{NR2} for the BRST charge
(\ref{4.13}), (\ref{4.33}) from Eq.(\ref{4.15}) we obtain  that
\begin{eqnarray}
H_A &=& H + p^\alpha \Bigl(
\u{0}^\beta_\alpha \Bigr)^0\,\p{1}_\beta + \sum_{k=1}^N \th{k}^\alpha \left(
\p{k+1}_\alpha - \left( \u{k}^\beta_\alpha + \mu^\delta
\A{1}{N-k+1}^\beta_{\alpha \delta} \right)^0\,\p{1}_\beta \right) \nonumber \\
&& \nonumber \\
&+& p^\alpha \left(\frac{\partial H}{\partial p_r} \chi^\beta_{rs}
\frac{\partial \Ph{0}_\alpha}{\partial p_s}\right) \bar{c}_\beta +
\sum_{k=1}^N \th{k}^\alpha \left( \frac{\partial H}{\partial p_r}
\chi^\beta_{rs} \frac{\partial \Ph{k}_\alpha}{\partial p_s} \right)
\bar{c}_\beta + \Delta{H_A} , \label{4.45}
\end{eqnarray}
where
$\Delta{H_A}$ contains the terms having more than quadratic powers in
the ghost variables. In this, the Hamiltonian $H$ and all the
structure functions in Eq.(\ref{4.45}) are the standard functions.
Comparing Eqs.(\ref{3.39}) and (\ref{4.45}), using
the identification rules (\ref{4.35}), (\ref{4.36}) and taking into
account the arbitrariness in definition of the Hamiltonian and
constraints we see that one can always construct the BRST invariant
Hamiltonian within the framework of BFV formalism in such a way, that
it will coincide with the Hamiltonian, obtained from the effective
BRST invariant Lagrangian.

\section{Conclusion}

We have constructed both Lagrangian and Hamiltonian BRST formalisms for the
systems having the gauge symmetry under transformations (\ref{2.1}), forming
closed gauge algebra, and proved the equivalence of these two
approaches.  In this way, we have also obtained an explicit relation
between the so-called gauge fermions, which remove the degeneracy of
the system within the framework of Lagrangian and Hamiltonian BRST
formalisms.  Besides, having used the Ostrogradsky formalism we shown the
correspondence between the ghosts with higher order time derivatives of
the Lagrangian approach and the canonical ghosts of the BFV formalism.
Note that the Lagrangian BRST charge written in the terms of the Hamiltonian
variables has been expressed through the standard constraints.
We observed the same appearence of the standard extension in
Refs.\cite{NR1,NR2,NR3}, where the case of $N = 1$ with both closed
and open gauge algebras had been considered.

Obviously, to simplify the calculations it is desirable to suppose
the quantities $\chi^\alpha_{rs}$ from Eq.(\ref{3.36}) to be equal
to zero for any values of the velocity phase space coordinates
$q^r, \dot q^r$ (i.~e. globally). It can be shown \cite{NR1} that
there exists the corresponding choice of the vectors $\chi^\alpha_r$
in general if, and only if, the vector fields
$\u{0}^\beta_\alpha \ps{0}^r_\beta {\partial}/{\partial q^r}$
form an abelian Lie subalgebra of the gauge algebra.

Note finally that our consideration, done on the classical level, allows us
to prove the equivalence between Lagrangian and Hamiltonian BRST formalisms
on quantum level as well.

\vskip0.5cm

The author is grateful to Profs. A.V. Razumov, V.A. Rubakov and
F.V. Tkachov for support, useful discussions and valuable remarks.
This research was supported in part by the
International Science Foundation under grant  MP 9000 .


\begin{thebibliography}{**}

\small

\bibitem{BRST}
C. Becchi, A. Rouet and R. Stora, Phys. Lett. B52 (1974) 344; Commun. Math.
Phys. 42 (1975) 127; Ann. Phys. 98 (1976) 287; \\
I.V. Tyutin, FIAN preprint 39 (1975)

\bibitem{Lagr}
T. Kugo and S. Uehara, Nucl. Phys. B197 (1982) 378;\\
F.R. Ore and P. van Nieuwenhuizen, Nucl. Phys. B204 (1982) 317; \\
L. Alvarez--Gaum\'e and L. Baulieu, Nucl. Phys. B212 (1983) 255; \\
L. Baulieu, Phys. Rep. 129 (1985) 1

\bibitem{BFV}
E.S. Fradkin and G.A. Vilkovisky, Phys. Lett. B55 (1975) 224; \\
I.A. Batalin and G.A. Vilkovisky, Phys. Lett. B69 (1977) 309; \\
E.S. Fradkin and T.E. Fradkina, Phys. Lett. B72 (1978) 343; \\
I.A. Batalin and E.S. Fradkin, Phys. Lett. B122 (1983) 157; \\
I.A. Batalin and E.S. Fradkin, Riv. Nuovo Cim. 9(10) (1986) 1

\bibitem{Hen}
M. Henneaux, Phys. Rep. 126 (1985) 1

\bibitem{Equiv}
C. Battle, J. Gomis, J. Par\'\i s and J. Roca, Pys. Lett. B224 (1989) 288;
Nucl. Phys. B239 (1990) 139; \\
J.M.L. Fisch and M. Henneaux, Phys. Lett. B226 (1989) 80; \\
W. Siegel, Int. J. Mod. Phys. A4 (1989) 3705;
Int. J. Mod. Phys. A4 (1989) 3953; \\
G.V. Grigoryan, R.P. Grigoryan and I.V. Tyutin, Sov. J. of Nucl. Phys. 53(6)
(1991) 1720

\bibitem{NR1}
Kh.S. Nirov and A.V. Razumov, BRST formalism for systems of Yang--Mills type,
IHEP preprint 90-45 (Protvino, 1990)

\bibitem{NR2}
Kh.S. Nirov and A.V. Razumov, J. Math. Phys. 34 (1993) 3933

\bibitem{NR3}
Kh.S. Nirov and A.V. Razumov, Int. J. Mod. Phys. A7 (1992) 5719

\bibitem{N}
Kh.S. Nirov, Constraint algebras in gauge invariant systems,
INR preprint--860/94 (Moscow, 1994)

\bibitem{Arn}
V.I. Arnold, Mathematical methods of classical mechanics,
(Nauka, Moscow, 1989, in Russian)

\bibitem{PR}
P.N. Pyatov and A.V. Razumov, Int. J. Mod. Phys. A4 (1989) 3211

\bibitem{NPR}
Kh.S. Nirov, P.N. Pyatov and A.V. Razumov, Int. J. Mod. Phys. A7 (1992) 5549

\bibitem{DeW}
B. DeWitt, Supermanifolds (Cambridge University Press, Cambridge, 1984)

\bibitem{PRR}
P.N. Pyatov, A.V. Razumov and G.N. Rybkin, Classical mechanics on superspace,
IHEP preprint 88-212 (Protvino, 1988)

\bibitem{Dir}
P.A.M. Dirac, Lectures on quantum mechanics (Yeshiva University, New-York,
1964)

\bibitem{RS}
A.V. Razumov and L.D. Soloviev, Introduction to classical mechanics of
constrained systems, IHEP preprints 86-212, 86-213, 86-214 (Protvino, 1986)

\bibitem{GT}
D.M. Gitman and I.V. Tyutin, Canonical quantization of fields with
constraints (Nauka, Moscow, 1986, in Russian)

\bibitem{R}
P. Lancaster, Theory of matrices (Academic Press, New-York, 1969); \\
A.V. Razumov, Dependent coordinates in classical mechanics, IHEP preprint
84-86 (Protvino, 1984)

\end{thebibliography}
\end{document}